\documentstyle[12pt]{article}
\begin{document}

\title{ PROXIMITY  EFFECT  IN  BULK\ \  $LaBa_2Cu_3O_{7-y}$  SAMPLES  WITH\ \  $Ag$
ADDITIONS}

\author{E. Nazarova, A. Zahariev,  A. Angelow, K. Nenkov\\
Institute of Solid State Physics, 72 Trackia Blvd., 1784 Sofia, BG}

\maketitle

\begin{abstract}
Bulk $LaBa_2Cu_3O_{(7-y)}$ samples with different $Ag$ additions were
investigated. It was shown that $Ag$ does not enter the crystallographic
structure of the superconductor and segregates on the grain boundary region.
Current path in these samples occurs through the proximity connected grains
and this was confirmed from the temperature dependence of the critical current
density and mutual inductance. By using the theory developed for the thin
film structures we conclude that growing of the $Ag$ content increased the
effective cross section and the normal metal thickness. The first one
prevails at low concentration increasing the current. The second dominates at
higher concentration leading to saturation or even lowering of the critical
current.
\end{abstract}

\section{Introduction}\ \
Thin film structures prepared from low and high temperature superconducting
(HTSC) materials, based on the proximity effect, were widely discussed in
literature \cite{kn:simon}, \cite{kn:gijs},\cite{kn:deutscher} \cite{kn:durusoy}. Is it possible a
proximity coupling to accur between the grains through normal metal inclusions in bulk
HTSC samples? It is difficult to answer to this question having in mind low
value and high anysotropy of coherence length ($\xi$) in HTSC. Nevertheless
there were found some indications for the existence of the proximity
effect in bulk samples \cite{kn:lubberts}, \cite{kn:pinto}, \cite{kn:jung}.
The more frequently used experiments proving the existence of proximity
junction like I-V and $dI/dV$ curves, temperature dependence of normal layer
coherence length ($\xi_n$) or exponential decay of $I_c$ with increasing of
the normal layer thickness, L, are not applicable to the case of bulk samples.
The aim of this work is to indicate that the critical
current in bulk samples with normal metal inclusions can be governed by the
proximity effect. This is very important from practical point of view because
similar conditions can exist in Ag-sheathed tapes and wires.

\newpage
\section{Experimental}\ \

The investigated samples were prepared by the solid state reaction method
described in detail previously \cite{kn:nazarova}. Five samples with different
$Ag$  additions (2 wt\%, 5 wt\%, 10 wt\%, 15 wt\% and 20 wt\%) and one
without $Ag$ additions were examined.
The specimens were characterized by X-ray diffraction analysis using DRON 4M
powder diffractometer with $Cu_{k\alpha}$ radiation and Scanning Electron
Microscope Philips 515 with EDX analysis. Temperature dependence of critical
current at zero magnetic field was measured in a specially designed device.
The necessary temperature was maintained by thermoregulator within $\pm 0.15$ K
with an accuracy of $\pm 0.01$ K.
The appearance of $10 \mu v/cm$ voltage is accepted as a critical current
criterion.
The "screening" method (according to \cite{kn:claassen}) was used for the
mutual inductance determination. \ \

\section{Results and discussions}

$LaBa_2Cu_3O_{7-y}$ samples with starting composition 1:2:3 usually contain a
small amount of $BaCuO_2$ impurity phase \cite{kn:wada}, \cite{kn:song}.
X-ray diffraction analysis
of our undoped samples also detected the presence of $BaCuO_2$. The impurity
phase is not observed for the $Ag$ doped samples.
A very small shift (less than 0.05 degree)
of the main orthorhombic peaks was observed for the $Ag$ doped samples.
This shift corresponds to the change of $"c"$ lattice parameter from 11.8329 A
for undoped sample to 11.8256 A for 20 wt\% $Ag$ doped sample.
Such deviation, however, cannot be due to $Ag$ incorporation in 1:2:3
unit sell because it is shortenning and  very small.
$Ag$ is melted during
the process of sample synthesis and incorporates significant amount of oxygen.
It is more reasonable to think that the inclusion of this oxygen in 1:2:3 phase
is able to
produce the observed reduction in $"c"$ lattice parameter.\ \

SEM investigations also support the $Ag$ segregation. EDX analysis in the
superconducting grains shows nominal compositions close to
1:2:3 stoihiometry and no $Ag$ within detectable
limits has been found. $Ag$ additions are statistically
uniformly distributed through the specimen between the grains.\ \

Therefore we assume that $Ag$ inclusions do not enter the crystallographic structure
of $LaBa_2Cu_3O_{7-y}$. During the synthesis process $Ag$ moves out of growing
grains and
segregates on the grain boundary regions. Similar results were observed for
the $Ag$ doped $YBCO$ samples \cite{kn:singh}, \cite{kn:lin} and $Bi-2223$
phase \cite{kn:kawasaki}.\ \

For doped and undoped samples different grain boundary conditions exist.
The variations of the grain boundary content significantly influences
the critical current of the samples. Grain boundaries of undoped
samples mainly contain the insulating impurity phase. It is reasonable to
expect that current path occurs through the superconductor -- insulator --
supeconductor (S-I-S) system. In doped samples, with the increasing of the $Ag$
content, the current flows predominantly through the $Ag$ coupled grains. These
samples can be considered as a superconductor -- normal metal -- superconductor
(S-N-S) system on the level of individual grains. In case of individual
junction the $I_c(T)$ dependence over temperature range below $T_c$
provides a clear inside into the nature of the junction. This dependence
can be used to distinguish $S-N-S$ devices from other junction types. \ \

In our samples we have normal metal barriers
with different thickness and electron mean free path is not limited
from the thickness of the normal interlayer as in the thin film structures.
The average dimension of the $Ag$
particles determined from the width of the $Ag$ x - ray diffraction pick is
of order of 30 nm for the sample with 20 wt\%.
The normal metal coherent length $\xi_n$ determines the mean size of the Cooper
pair in the normal metal barier.

Assuming that the carrier mean free path $l_n$ in the normal
metal is determined from the dimensions of the $Ag$ particles both limits:
"clean"  ($l_n > \xi_n$) and "dirty" ($l_n < \xi_n$) will be present at
nitrogen temperatures in our unusual for the proximity studies samples.
But only $S-N-S$ junctions in "dirty" limit will support the transport current
fascilitating the grains connection in the specimens. In fact normal metal thickness,
L, has to be of order of $\xi_n$ for achieving a intergranular current close to
intragranular one.
By decreasing the temperature, T, normal metal cocherence length
will increase according to the relation \cite{kn:delin}:
\begin{equation}\label{eq:xx}
\xi_n = \left ( {{\hbar v_n l_n} \over {6\pi kT}} \right )^{1 \over 2},
\end{equation}
where $v_n$ is the Fermi velocity of normal metal and $k$ is the Boltzman
constant. At nitrogen temperatures $\xi_n$ will be about 15 nm
and will increase up to 66 nm at 4 K for $l_n$ equal to the
average partical size of $Ag$.
Some of the $S-N-S$ transitions being in "clean" limit at nitrogen temperature
will turn to "dirty" limit at lower temperatures. This will increase to some
extent the number of $S-N-S$ connections able to support high intergranular
current.\ \

The systematic theoretical investigation of the behavior of $S-N-S$ thin film
structures was carried out first by De Gennes\cite{kn:degennes}.
An expression has been derived for $I_c$ as a function of temperature T
and L:
\begin{equation}\label{eq:jctl}
{I_c(T,L)} = {{\pi A|\Delta_i|^2 L} \over {2e R_n kT_c \xi_n}}\exp \left(-L \over \xi_n \right),
\end{equation}
where $\Delta_i$ is the superconducting gap of the normal interlayer
interface, $A$ is the cross section of the individual $S-N-S$ transition, $R_n$
is the normal layer resistance, $T_c$ is the superconductor transition
temperature and $e$ is the electron charge. Eq. (\ref{eq:jctl})
was derived using dirty limit ($l_n < \xi_n$) boundary conditions. \ \

Following Delin at all \cite{kn:delin} for an $S-N-S$ structure prepared
from a high --$T_c$ superconductor and a noble metal as interlayer, $\Delta_i^2$ will have the form:
\begin{equation}\label{eq:delta}
\Delta_i^2 = \Delta^2{2 \over \pi^2 \gamma^2}\ {{T_c - T} \over T},
\end{equation}
where $\Delta$ is the superconductor energy gap far from the interface
region and $\gamma$ is the so called interface parameter.
The value of $\gamma$ is given by the ratio
\begin{equation}\label{eq:gamma}
\gamma = \left( {N_n \rho_s} \over {N_s \rho_n} \right)^{1 \over 2},
\end{equation}
where $N_n$ and $N_s$ are the normal metal and superconductor
density of states, respectively,
and $\rho_n$ and $\rho_s$ are their resistivities, respectively.
For high -- $T_c$ superconductor -- noble metal ($Ag$) junction $N_n \gg N_s$
($N_n = 5.85.10^{28} m^{-3}$ and for YBCO $N_s = 5.0.10^{27} m^{-3}$) and
$\rho_n \ll \rho_s$ ($\rho_n \approx 0.0084.10^{-6} \Omega .m$ and for YBCO
$\rho_s = 0.77.10^{-6} \Omega .m$ at 77K) therefore $\gamma \gg 1$
($\gamma \approx 1078$). \ \

Substituting Eq.(\ref{eq:xx}) and Eq.(\ref{eq:delta}) in Eq.(\ref{eq:jctl})
and assuming a BCS temperature dependence of the energy gap $\Delta (T)$  in the HTSC material:
\begin{equation}\label{eq:delta1}
\Delta (T) \approx 3.2\ k T_c\ \left( 1- {T \over T_c} \right)^{1\over 2},
\end{equation}
an useful expression for $I_c(T,L)$ is obtained:
\begin{equation}\label{eq:jctl1}
I_c(T,L) =
{{(3.2)^2 \sqrt 6 k^{3 \over 2} A T^{1 \over 2}{(T_c - T)^2}} \over {e \rho_n  \gamma^2 (\pi \hbar v_n l_n)^{1 \over 2}}}
\exp \left[ -L \left (6 \pi k T \over \hbar v_n l_n \right)^{1 \over 2} \right].
\end{equation}

Eq.(\ref{eq:jctl1}) present the temperature dependence of the critical
current across the $S-N-S$ structure. We will use this equation to explain
our experimental results.
Dividing both sides of Eq.(\ref{eq:jctl1}) by the sample cross
section, in the left side we obtain the critical current density through the
sample. It is seen that temperature dependence of the critical current density
is following:
\begin{equation}\label{eq:jt}
J_c(T) \propto T^{1 \over 2}(T_c-T)^2 \exp (-T^{1 \over 2}).
\end{equation}
The term $(T_c - T)^2$ will dominate this
dependence because the exponential and $T^{1 \over 2}$ terms have a weak influence.
This is shown in the Fig.1. It present $J_c(T)$ experimental  dependencies for the
$Ag$ doped samples. The experimental data for sample with 20 wt\% $Ag$ are
fitted twice: fit 1 - by the $(T_c - T)^2$ term only, and fit 2 - by the
temperature dependence according to (\ref{eq:jt}). Both fits show  very close
results which give us reason to use the simple one ignoring in fact $\xi_n(T)$.
It was found also that the quadratic fit (straight lines) gives the  least
mean square deviation from the data for 5 to 20 wt\% $Ag$ doped samples,
when the fits of type $(T_c - T)^n$ with $n \leq 2$ were examined.\ \

For the sample with 2 wt\% $Ag$ experimental results can  be fitted
better with linear temperature dependence. The quadratic fit is also possible.
At this concentration it is difficult to create the current path entirely
through the $Ag$ connected grains. More frequently the connection occurs
through the insulating phase and this dependence resembles
$J_c$ vs. $(T_c - T)$ for undoped samples. For the comparison, on the Fig.2
$J_c$ vs. $(T_c - T)$ is
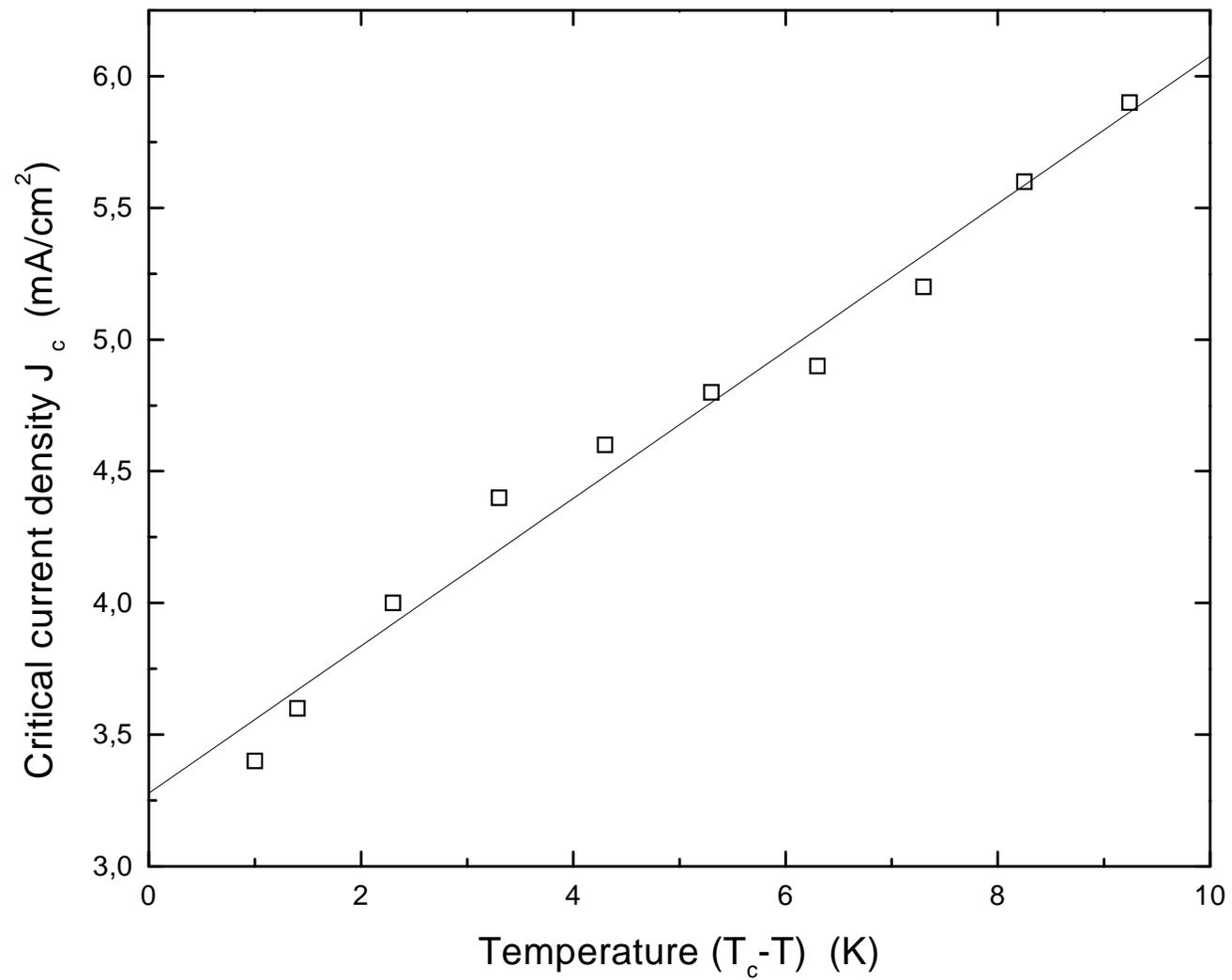
presented for undoped sample. The line gives the best linear fit to the data.
The temperature dependence of current across the $S-I-S$ Jousephson junction
is described by the Ambegaokar - Baratoff dependence \cite{kn:ambegaokar}.
For values $T < T_c$ but close to $T_c$  a linear temperature dependence occurs.
By lowering the temperature the current reachs the saturation. In fact $Ag$
additions change
the mechanism of current flow on the basis of  the proximity effect and
its value increased. \ \

In the Fig.3 $\sqrt{J_c}$ vs. $(T_c -T)$ plots are shown for $Ag$ doped
samples.
It is seen that with increasing of the $Ag$ content the slopes of the lines
increase and a tendency to a saturation can be noticed too. According to Eq.(\ref{eq:jctl1})
$A$ and $L$ quantities can be influenced by the $Ag$ amount. For our samples
$A$ has a meaning of an effective cross section. With increasing of the $Ag$
amount both quantites $A$ and $L$ could increase too.
At small $Ag$ concentrations the firs one prevails raising the value of
critical current density in the sample. At higher concentration $A$ goes to
a saturation and $L$ -- increasing begins to dominate, lowering the probability for
the proximity connection between the grains. This leads first to
a saturation and then to a current decreasing when $L> (2-3) \xi_n$.
This can serve as a basis of explanation of the observed saturation tendency in our case
and the
lowering of the critical current density reported by some authors \cite{kn:jung}, \cite{kn:singh},
\cite{kn:itoh} when the $Ag$ additions come nearer to some critical
concentration. This concentration is technologically dependent.\ \

In the Fig.4 normalyzed mutual inductance is presented as a function of
temperature for the non doped sample and sample with 20 wt\% $Ag$ additions.
It is seen that in wide temperature interval below $T_c$ the sample with $Ag$ has
a lower mutual inductance than non doped. As the mutual inductance is
proportional to the density of superconducting electrons, $n_s$\cite{kn:claassen}.
it is lowered in the sample with $Ag$ due to the lowering of $n_s$ in normal metal
inclusions. $Ag$ particles with different thickness $L$ exsit in the sample. By
lowering the temperature below $T_c$, they will be able to connect superconducting
grains when their thickness $L$ becomes of order of $\xi_n$ according
to Eq.(\ref{eq:xx}).\ \

These investigations are important from the point of view of practical
applications of HTSC materials in wires and tapes. For example the "powder
in tube" method, which is widely used for tapes preparation, ensure the existence
of HTSC materials/silver interface. During the heat treatment process of
the tape preparation, $Ag$ diffusion between the superconducting grains on
the interface
region is possible. Investigations of transport current distributions in $Ag$
sheathed $Bi-2223$ based tapes showed that high current densities are observed
in a thin layer ($(2-3) \mu m$ width) located close to the silver, which is roughly
10 - 15 \% of the total superconducting area \cite{kn:welp}. Therefore,
when the preparation technology ensures the
formation of $Ag$ particles with dimensions of order of $\xi_n$, the current
path through the specimens will be determined by the proximity effect.
Grain boundaries containing normal metal with various thickness can exist in
doped samples.  The current
path, however, will be through the boundaries with the thickness of order
of $\xi_n$.\ \

In conclusion we investigate bulk 1:2:3 HTSC material with normal metal ($Ag$)
inclusions with an average dimension of order of $\xi_n$. It was shown that the
current path in these samples is based on the proximity coupling between the
grains. We establish that $J_c(T)$ dependence below $T_c$ is quadratic
which is typical for $S-N-S$ structures. Mutual inductance measurements also
support this type of intergranular current.
By using the theory developed for thin film structures we analyze the
experimental results for the samples with different $Ag$ amount.
When the $Ag$ additions increased the number of proximity coupled grains and
normal metal thickness increased, too. If the first one dominates the critical
current in the sample will increase, while the domination of the second leads to
a saturation or current decreasing when $L> (2-3) \xi_n$. The value of the current
through connected in series $S-N-S$ transitions is limited by the grains
with the smallest cross section, A, and the largest thickness, L.

\newpage

\section{Acknowledgment}
The authors are grateful to Dr. Ram Kossowsky supporting the presentation
of the work at the NATO ARW.

This work was partly supported by the National Foundation
"Scientific Research" under Grant 421.\ \

\end{document}